\documentclass[preprint,pre,showpacs]{revtex4}
\usepackage{graphicx}% Include figure files
\begin{document}
\title{Noise in Disordered Systems: Higher Order Spectra in Avalanche Models } 
\author{ Amit P. Mehta }
\email[]{apmehta@uiuc.edu}
\affiliation{ Department of Physics, University of Illinois at
Urbana-Champaign, 1110 West Green Street, Urbana, IL 61801-3080 }
\author{Karin A. Dahmen}
\email[]{dahmen@uiuc.edu}
\affiliation{ Department of Physics, University of Illinois at
Urbana-Champaign, 1110 West Green Street, Urbana, IL 61801-3080 }
\author{A. C. Mills}
\affiliation{ Department of Physics, University of Illinois at
Urbana-Champaign, 1110 West Green Street, Urbana, IL 61801-3080 }
\author{M. B. Weissman}
\email[]{mbw@uiuc.edu}
\affiliation{ Department of Physics, University of Illinois at
Urbana-Champaign, 1110 West Green Street, Urbana, IL 61801-3080 }

\date{\today} 

\newcommand{\summ}{\sum_{m=1}^{j-1}}
\newcommand{\be}{\begin{equation}}
\newcommand{\ee}{\end{equation}}
\newcommand{\bea}{\begin{eqnarray*}}
\newcommand{\eea}{\end{eqnarray*}}
\newcommand{\la}{\langle}
\newcommand{\ra}{\rangle}
\newcommand{\xp}{\tilde{x}^+}
\newcommand{\yp}{\tilde{y}^+}
\newcommand{\xm}{\tilde{x}^-}
\newcommand{\ym}{\tilde{y}^-}
\newcommand{\yt}{Y_{2i}}
\newcommand{\ytp}{Y_{2i+1}}
\newcommand{\xt}{X_{2i}}
\newcommand{\xtp}{X_{2i+1}}
\newcommand{\mful}{\sum_{m=M/2}^{M-1}}
\newcommand{\mhaf}{\sum_{m=0}^{M/2-1}}
\newcommand{\nful}{\sum_{n=M/2}^{M-1}}
\newcommand{\nhaf}{\sum_{n=0}^{M/2-1}}
\newcommand{\lful}{\sum_{l=M/2}^{M-1}}
\newcommand{\lhaf}{\sum_{l=0}^{M/2-1}}
\newcommand{\jful}{\sum_{j=M/2}^{M-1}}
\newcommand{\jhaf}{\sum_{j=0}^{M/2-1}}
\newcommand{\kful}{\sum_{k=M/2}^{M-1}}
\newcommand{\khaf}{\sum_{k=0}^{M/2-1}}
\newcommand{\qful}{\sum_{q=M/2}^{M-1}}
\newcommand{\qhaf}{\sum_{q=0}^{M/2-1}}

\begin{abstract}

        Utilizing the Haar transform, we study the higher order
        spectral properties of mean field avalanche models, whose
        avalanche dynamics are described by Poisson statistics at a
        critical point or critical depinning transition. The Haar
        transform allows us to obtain a time series of noise powers,
        $H(f_1,t)$, that gives improved time resolution over the
        Fourier transform.  Using $H(f_1,t)$ we analytically calculate
        the Haar power spectrum, the real $1.5$ spectra, the second
        spectra, and the real cross second spectra in mean field
        avalanche models. We verify our theoretical results with the
        numerical results from a simulation of the $T=0$ mean field
        nonequilibrium random field Ising model (RFIM).  We also
        extend our higher order spectra calculation to data obtained
        from a numerical simulation of the $T=0$ infinite range RFIM
        for $d=3$, and experimental data obtained from an amorphous
        alloy, $Fe_{21}Co_{64}B_{15}$. We compare the results and
        obtain novel exponents.
        
\end{abstract}


\maketitle
%\vskip2pc]
%\vskip2pc
\section{Introduction} 

There are disordered systems that respond to slow driving with
discrete jumps or avalanches with a broad range of sizes, referred to
as {\em crackling} \cite{Nature}. Such crackling systems are
characterized by many interacting degrees of freedom and strong
interactions that make thermal effects negligible, such as: charge
density waves, vortices in type II superconductors, crack propagation,
earthquakes, and Barkhausen noise in magnets.  The avalanche dynamics
of the systems mentioned above are described by Poisson statistics
(given by Eq.  (\ref{prodist})) in mean field theory at a critical
point or critical depinning transition \cite{Matt1,Fisher}.

Through our analysis of mean field avalanche models we determine the
following spectral functions: Haar power spectrum, real $1.5$ spectra,
second spectra, and real cross second spectra for systems that have
avalanche dynamics given by Eq. (\ref{prodist}).  This analysis
provides new tools for noise analysis in dynamical systems, since
there are very few theoretical calculations of higher order noise
statistics.

        The Haar transform allows us to obtain a power versus time
        series, $H(f_1,t)$, needed to calculate higher order spectra.
        These higher order spectra give valuable information about the
        avalanche dynamics in Barkhausen noise not accessible through
        ordinary power spectra \cite{OBrien,Petta}.  Higher order
        spectra also have been used to obtain crucial information
        about a variety of diverse systems such as: metastable states
        in vortex flow \cite{Merithew}, natural auditory signals
        \cite{Thomson}, conductance-noise in amorphous semiconductors
        \cite{Parsam}, fluctuating current paths in devices
        \cite{Seidler}, and quasi-equilibrium dynamics of spin glasses
        \cite{Weissman}. While much experimental work has been done
        studying higher order spectra
        \cite{OBrien,Petta,Merithew,Thomson,Parsam,Seidler,Weissman},
        we present a rigorous mean field treatment that is applicable
        to a broad range of systems \cite{Fisher}. This analysis will
        allow a better understanding of the dynamics of these systems,
        and provide a direct method of comparison to experiment or
        observation.
        
        In addition, we compare our general results from mean field
        theory to Barkhausen noise obtained: from a mean field
        simulation of the $T=0$ random field Ising model (RFIM)
        \cite{Dahmen}, from a simulation of the $T=0$ infinite range
        RFIM (IRM) in $d=3$, and from experiment. We also find novel
        exponents from our analysis, and we compare our results from
        theory, simulation, and experiment; we find key similarities and
        differences.

\section{THE MODEL}
\subsection{Mean Field RFIM}

The $T=0$ mean field RFIM consists of an array of $N$ spins ($s_i =
\pm 1$), which may point up ($s_i = +1$) or down ($s_i = -1$). Spins
are coupled to all other spins (through a ferromagnetic exchange
interaction $J$), and to an external field $H(t)$ which is increased
adiabatically slowly.  To model disorder in the material, we assign a
random field, $h_i$, to each spin, chosen from a distribution $P(h_i)
= exp(-h^2_i/2R^2)/\sqrt{2\pi}R$, where $R$ determines the width of
the Gaussian probability distribution and therefore gives a measure of
the amount of quenched disorder for the system. The Hamiltonian for the
system at a time $t$ is given by: $H = -\sum\limits_i(JM + H(t) +
h_i)s_i$, where $M = \frac{1}{N}\sum\limits_j s_j$ is the
magnetization of the system. Initially, $H(-\infty) = -\infty$ and all
the spins are pointing down. Each spin is always aligned with its
local effective field $h_i^{eff} = JM + H(t) + h_{i}$.

\subsection{RFIM with infinite range Forces (IRM) in 3D}

        The $T=0$ Infinite
        Range RFIM (IRM) consists of a 3D lattice of $N$ spins, with a
        Hamiltonian at a time $t$ given by: $H = \sum\limits_{<ij>}
        -J s_i s_j - \sum_{i}(H(t) + h_i - J_{inf}M)s_i$, where
        $J_{inf}>0$ is the strength of the infinite range demagnetizing
        field, and $\langle ij\rangle$ stands for nearest neighbor
        pairs of spins. The local effective field is given by
        $h_i^{eff} = J\sum_{<ij>}s_j + H(t) + h_{i} - J_{inf}M$.

        The addition of a weak $J_{inf} \sim \frac{1}{N}$ to the
        traditional RFIM causes the system to exhibit self-organized
        criticality (SOC) \cite{Urbach,Narayan,Zapperi}.  This means
        that as $H$ is increased the model always operates at the
        critical depinning point, and no parameters need to be tuned
        to exhibit critical scaling behavior (except
        $\frac{dH}{dt}\rightarrow 0$).  We limit our analysis to a
        window of $H$ values where the slope of $M(H)$ is constant and
        the system displays front propagation behavior.  Details of
        the simulation algorithm are given elsewhere \cite{Matt2}.

\subsection{Avalanche Dynamics in the $T=0$ RFIM} 
        
        The external field $H(t)$ is adiabatically slowly increased
        from $-\infty$ until the local field, $h_i^{eff}$, of any spin
        $s_i$ changes sign, causing the spin to flip
        \cite{Matt1,Sethna}.  It takes some microscopic time $\Delta
        t$ for a spin to flip.  The spin flip changes the local field
        of the coupled spins and may cause them to flip as well, etc.
        This {\em avalanche} process continues until no more spin
        flips are triggered.  Each step of the avalanche, that is each
        $\Delta t$, in which a set of spins simultaneously flip, is
        called a {\em shell}. The number of spins that flip in a shell
        is directly proportional to the voltage $V(t)$ during the
        interval $\Delta t$ that an experimentalist would measure in a
        pick-up coil wound around the sample. In our simulations we
        denote the number of spins flipped in a shell at a time $t$ by
        $n_t (= V(t))$.  The first shell of an avalanche (one spin
        flip) is triggered by the external field $H(t)$, while each
        subsequent shell within the avalanche is triggered only by the
        previous shell, since $H(t)$ is kept constant while the
        avalanche is propagating.  $H(t)$ is only increased when the
        current avalanche has stopped, and is increased only until the
        next avalanche is triggered (i.e. $\frac{dH}{dt}\rightarrow
        0$).  The number of shells in an avalanche times $\Delta t$
        defines the {\em pulse duration}, $T$, or the time it took for
        the entire avalanche to flip.  The time series of $n_t$ values
        for many successive avalanches creates a Barkhausen train
        analogous to experiment.

\section{ EXPERIMENT }

We compare our results for theory and simulation with results obtained
from an experiment performed on an (unstressed) amorphous alloy,
$Fe_{21}Co_{64}B_{15}$. Measurements were performed on a 21 cm x 1 cm
x 30 $\mu$m ribbon of $Fe_{21}Co_{64}B_{15}$ alloy, a soft amorphous
ferromagnet obtained from Gianfranco Durin.  The domain walls run
parallel to the long axis of the material, with about 50 domains
across the width.  A solenoid, driven with a triangle wave, applies a
magnetic field along the long axis of the sample.  Since domain wall
motion dominates over other means of magnetization in the linear
region of the loop, data were collected in only a selected range of
applied fields near the center of the loop.  The Barkhausen noise was
measured by a small pick-up coil wound around the center of the
sample.  This voltage signal was amplified, anti-alias filtered and
digitized, with care taken to avoid pick-up from ambient fields.
Barkhausen noise was collected for both increasing and decreasing
fields for $80$ cycles of the applied field through a saturation
hysteresis loop.  The driving frequency was $0.01$ Hz; this
corresponds to $c=0.09$, where c is a dimensionless parameter
proportional to the applied field rate and is defined in the
Alessandro Beatrice Bertotti Montorsi model (ABBM model) for the
Barkhausen effect in metals \cite{ABBM}. In this way, our measurements
should be well inside the $c<1$ regime identified in the ABBM model,
in which we can expect to find more or less separable avalanches
rather than continuous domain wall motion.

\section{THEORY}
\subsection{Poisson Distribution}

The probability distribution for the avalanche dynamics in the class
of mean field avalanche models we are interested in is given by
\cite{Fisher,Matt1}:

\begin{equation} \label{prodist}
        P(n_0 = 1,n_1,n_2,...,n_{\infty}) = \frac{1}{e n_1!}
        \prod\limits^{\infty}_{t=2}\frac{e^{-n_{t-1}}n^{n_t}_{t-1}}{n_t!}
\end{equation}  

        The above probability distribution (Eq. \ref{prodist}) is for
        the time series of a {\em single} infinite avalanche at the critical
        point.  In the context of the mean field RFIM, $n_t$
        represents the number of spins flipped at a time $t$. That is,
        each $n_t$ represents a shell of the avalanche, and since an
        avalanche begins with one spin flip we have that $n_0 = 1$.
        
        Let $\la .\ra$ represent the average over Eq. \ref{prodist}.
        In order to calculate the Haar transform and higher order
        spectra in mean field theory we need the following quantities,
        where $m\ge l\ge k\ge 0$ and $j\ge 0$:

\begin{eqnarray}\label{avg1}
        && \la n_j\ra = 1\\ 
\label{avg2}
        && \la n_j^2\ra = j+1\\
\label{avg3}
        && \la n_jn_{j+k}\ra = \la n_j^2\ra = j+1\\
\label{avg4}
        && \la n_j^3\ra = \frac{1}{2}(3j^3+5j+2)\\
\label{avg5}
        && \la n_j^4\ra = \frac{1}{2}(6j^3+13j^2+9j+2)\\ 
\label{avg6}
        && \la n_jn_{j+k}n_{j+l}\ra = \la n_j^3\ra + k\la n_j^2\ra\\
\label{avg7}
        && \la n_jn_{j+k}n_{j+l}n_{j+m}\ra\nonumber\\
        && = \la n_j^4\ra+(2k+l)\la n_j^3\ra+\frac{k}{2}(k+2l-1)\la n_j^2\ra 
\end{eqnarray}

These relations can be determined from Eq. (\ref{prodist}). We notice
that the time ordering of the indices plays on important role.
Details of how Eqs. (\ref{avg1}-\ref{avg7}) were derived are given
in Appendix \ref{corcalc}.

\subsection{Haar Power} 

The Haar transform is a simple wavelet transform with basis states
consisting of single-cycle square waves. We use the Haar transform
rather than the Fourier transform since it gives us improved time
resolution in exchange for less frequency resolution. Time resolution
is important to our purpose since we are interested in studying how
the power contribution around a frequency $f_1$ changes along the
duration of the avalanche.
        
Physically, the Haar power, $H(t,f_1)$, is the absolute square of the
time integral over a period (of duration $1/f_1$) of a
single-cycle square wave times a section of the train centered around $t$
\cite{OBrien}. In other words, to determine the Haar power we integrate a
square wave of duration $1/f_1$ over a section of the noise
train centered at time $t$; this integrated segment is then squared to
assure our resulting values is positive definite, this squared segment
corresponds to $H(t,f_1)$, the Haar power at time $t$ and around
frequency $f_1$.  In order to analytically determine the Haar power
series we first define the sum over $N$ shells:

\begin{eqnarray}\label{sum}
        && \tilde{x}_i^+(N) = \sum_{l=0}^{N/2-1}n_{Ni+l}\Delta t\\
        && \tilde{x}_i^-(N) = \sum_{l=N/2}^{N-1}n_{Ni+l}\Delta t
\end{eqnarray}

Where $t = Ni$, and $\Delta t$ is the time separation between shells.
The Haar power series for a frequency $f_1 =
\frac{1}{N\Delta t}$ is defined as:

\begin{equation}\label{haar1}
        H(f_1=\frac{1}{N\Delta t},i=\frac{t}{N}) \equiv \la (\tilde{x}_i^+(N)-\tilde{x}_i^-(N))^2\ra
\end{equation}

        To evaluate Eq. \ref{haar1} we determine the following
        relations:

\begin{eqnarray}\label{avgsum1}
        && \la \tilde{x}_i^\pm(N)\ra = N\Delta t/2\\ 
\label{avgsum2}
        && \la (\tilde{x}_i^+(N))^2\ra =
        \sum_{m=0}^{N/2-1}[2(N/2-1-m)+1]\la n^2_{Ni+m}\ra\\
\label{avgsum3}
        && \la (\tilde{x}_i^-(N))^2\ra =
        \sum_{m=N/2}^{N-1}[2(N-1-m)+1]\la n^2_{Ni+m}\ra\\
\label{avgsum4}
        && \la \tilde{x}_i^+(N)\tilde{x}_i^-(N)\ra 
           = \frac{N}{2}\sum_{m=0}^{N/2-1}\la n^2_{Ni+m}\ra
\end{eqnarray}

        When evaluating Eq. (\ref{avgsum2}-\ref{avgsum4}) we must take
        into account the time ordering of the indices, since the time
        ordering is needed to evaluate the ensemble average of 2-pt
        (Eq. \ref{avg3}), 3-pt (Eq. \ref{avg6}), and 4-pt
        (Eq. \ref{avg7}) correlation functions.  Refer to Appendix
        \ref{timeorder} for details.

        The above sums can be evaluated with the help of
        Eq. (\ref{avg2}).  Now using
        Eqs. (\ref{avgsum1})-(\ref{avgsum4}) we obtain the following
        exact result:

\begin{equation}\label{haar2}
        H(f_1=\frac{1}{N\Delta t},i) = \frac{1}{12}(\frac{(\Delta t)^2}{f_1^3}+\frac{1}{f_1}) 
\end{equation}

To find the Haar power spectrum we sum over all $i$, that corresponds
to the sum over the Haar wavelets.  To do this we define the maximum
duration of the avalanche to be $T$.  Now since blocks of $N$ shells
have been summed over, we perform a sum over Eq. (\ref{haar2}) from
$0$ up to $T/N-1$:

\begin{eqnarray}\label{haar3}
        && S_H(f_1) =
        \frac{1}{T}\sum\limits_{i=0}^{T/N-1}H(f_1=\frac{1}{N},i) \\
        && = \frac{1}{12}(2(\Delta t)^2 + \frac{1}{f_1^2}) \\
        && \simeq \frac{1}{12f_1^2}
\end{eqnarray}

The Haar power spectrum, $S_H(f_1)$, is in excellent agreement with
the Haar power spectrum determined from simulation. The Fourier power
spectrum, $S_F(f_1) = 1/f_1^2$ \cite{Matt1}, differs by an
additive constant and a constant factor from $S_H(f_1)$.  The additive
constant, which depends on $\Delta t$, is in effect a result of
aliasing produced by the discrete sampling.

\subsection{$1.5$ Spectra, Second Spectra, and Cross Second Spectra} 

        The $1.5$ spectra, second spectra, and cross second spectra
        are defined below:
        
\begin{eqnarray} \label{halfspec}
&& S_{1.5}(f_2,f_1) = \frac{\langle
F_t\{v(t,f_1)\}F_t^*\{H(t,f_1)\}\rangle}{\langle H(t,f_1)\rangle_t} \\
\label{secondspec}
&& S_2(f_2,f_1) = \frac{\langle
F_t\{H(t,f_1)\}F_t^*\{H(t,f_1)\}\rangle}{\langle H(t,f_1)\rangle_t^2} \\
\label{crosspec}
&& S_2(f_2,f_b,f_a) = \frac{\langle
F_{t_a}\{H(t_a,f_a)\}F_{t_b}^*\{H(t_b,f_b)\}\rangle}{\langle H(t_a,f_a)\rangle_{t_a}\langle
H(t_b,f_b)\rangle_{t_b}}
\end{eqnarray}

Where $v(t,f_1)$ is the sum of $n_i$ around the time $t$ over a
duration $1/f_1$, and $f_2$ is the frequency conjugate to $t$.
Also, $F_t\{...\}$ is the discrete Fourier transform with respect to
$t$. Now $\langle...\rangle_t$ denotes sum over $t$ over the duration
of the avalanche ($\sum_{t=0}^{T-1/f_1}$), similarly
$\langle...\rangle_{t_a} \equiv \sum_{t=0}^{T-1/f_a}$ and
$\langle...\rangle_{t_b} \equiv \sum_{t=0}^{T-1/f_b}$. For the cross
second spectra, $S_2(f_2,f_b,f_a)$, we require that $f_b > f_a$. In
addition, in the definition of $S_2(f_2,f_b,f_a)$ we require two
times, $t_a$ and $t_b$. The time $t_a$ labels the starting point of a
single-cycle square wave with a period of $1/f_a$ along the Barkhausen
train, and therefore $t_a$ takes on values that are multiples of
$1/f_a$. Similarly, $t_b$ labels the starting point of a single-cycle
square wave with a period of $1/f_b$, and takes on values that are
multiples of $1/f_b$.

To calculate Eq. (\ref{halfspec}-\ref{crosspec}) we first write the
product of Fourier transforms as the Fourier transform of a
convolution.  This mathematical identity allows us to separate the
Fourier transform ($F_t\{...\}$) from the ensemble average ($\la
.\ra$) This leaves us with the following quantities: $\la
v(t,f_1)H(t+\theta,f_1)\ra$, $\la H(t,f_1)H(t+\theta,f_1)\ra$, and
$\la H(t_a+\theta,f_a)H(t_b,f_b)\ra$ where $\theta$ is the convolution
variable.  These quantities may then be rewritten as a sum of 3-pt or
4-pt correlation functions, and subsequently evaluated.  We obtain the
general scaling forms:

\begin{eqnarray} \label{theta1}
&&\Gamma_{1.5}(f_1)\equiv \sum_{t=0}^{T-1/f_1}\langle v(t,f_1)H(t,f_1)\rangle \simeq
\frac{A_{1.5}}{f_1^{Q_{1.5}}}\\
\label{theta2}
&&\Gamma_2(f_1)\equiv \sum_{t=0}^{T-1/f_1}\langle H(t,f_1)H(t,f_1)\rangle \simeq
        \frac{A_2}{f_1^{Q_2}}\\
\label{theta3SC}
&&\Gamma_2(f_b,r) \equiv r\sum\limits_{t_a=0}^{T-1/f_a}
\sum\limits_{t_b=t_a}^{t_a+1/f_a-1/f_b} \langle
H(t_b,f_b)H(t_a,f_a)\rangle \simeq \frac{D_2 r^3}{f_b^{Q_2}} \\
\label{halfSC}
&& Re\{S_{1.5}(f_2,f_1)\} = \frac{B_{1.5}}{f_2^{V_{1.5}}}+ \Gamma_{1.5}^{(N)}(f_1)\\
\label{secondSC}
&& S_2(f_2,f_1) =  \frac{B_2}{f_2^{V_2}} + \Gamma_2^{(N)}(f_1)\\
\label{crossSC}
&& Re\{S_2(f_2,f_a,f_b)\} = \frac{B_2 r}{f_2^{V_C}} + \Gamma_2^{(N)}(f_b,r)
\end{eqnarray}

\begin{figure}
\includegraphics[scale= .45]{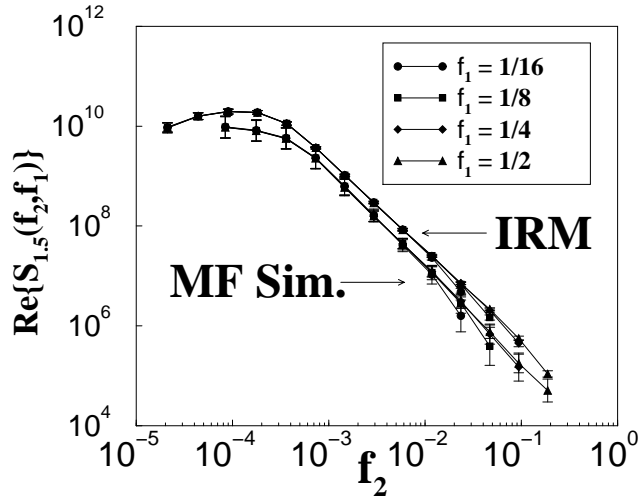}% Here is how to import EPS art
\caption{\label{halfspecplot} We present the $Re\{S_{1.5}(f_2,f_1)\}$ in the MF
simulation and in the IRM. At high frequency $Re\{S_{1.5}(f_2,f_1)\}\sim
f_2^{-V_{1.5}}$ (go to Table \ref{expo}). There is not a visible
flattening present due to $\Gamma_{1.5}^{(N)}(f_1)$ ($\theta=0$) term,
since the magnitude of this term is small relative to the $f_2$ dependent
term. }
\end{figure}

\begin{figure}
\includegraphics[scale= .45]{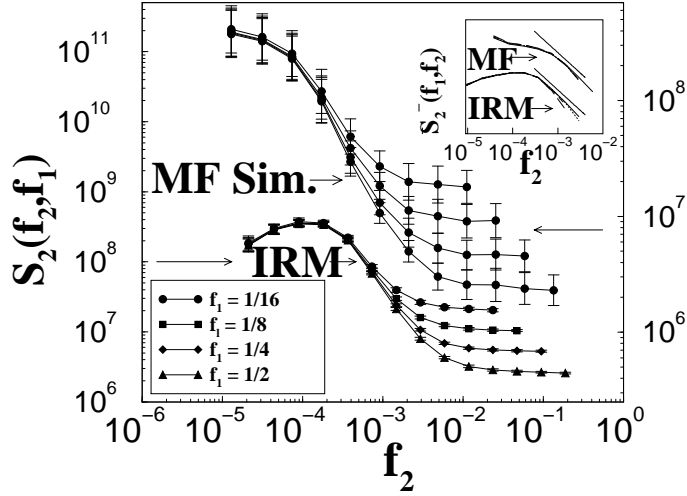}% Here is how to import EPS art
\caption{\label{secondspecplot} We present the $S_2(f_2,f_1)$ in the MF
and IRM. Notice the flattening due to the $\Gamma_2^{(N)}(f_1)$
($\theta=0$) term. Inset: $S_2^-(f_2,f_1)$ is the second spectra in the MF
simulation and the IRM with background term subtracted. The
bold lines adjacent to the MF curve is a power law with an exponent
of -2, and the bold line adjacent to the IRM curve has an exponent of
-1.8.}
\end{figure}

\begin{figure}
\includegraphics[scale= .45]{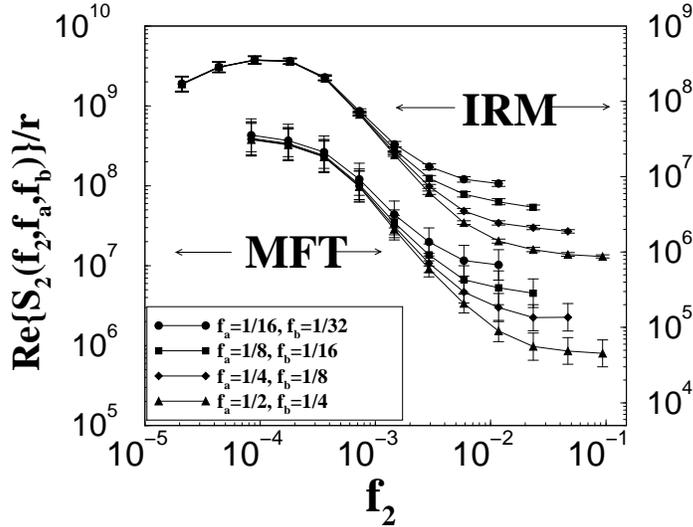}% Here is how to import EPS art
\caption{\label{crosspecplot} We present the $Re\{S_2(f_2,f_a,f_b)\}$
for $r= f_a/f_b = \frac{1}{2}$ in the MF simulation and the
IRM. Notice the flattening due to the $\Gamma_2^{(N)}(f_b,r=\frac{1}{2})$
($\theta=0$) term. }
\end{figure}

The exponents $V_{1.5}$, $V_2$, $Q_{1.5}$, $Q_2$ are given in Table
\ref{expo} for mean field theory, the IRM, and the experiment. Also,
$A_{1.5}$, $A_2$, $B_{1.5}$, and $B_2$ are non-universal constants. In
Eq. (\ref{theta3SC}) and Eq. (\ref{crossSC}) $r=f_a/f_b$. Plots of Eqs.
(\ref{halfSC}-\ref{crossSC}) are given in Figs.
(\ref{halfspecplot})-(\ref{crosspecplot}).

\begin{figure}
\includegraphics[scale= .45]{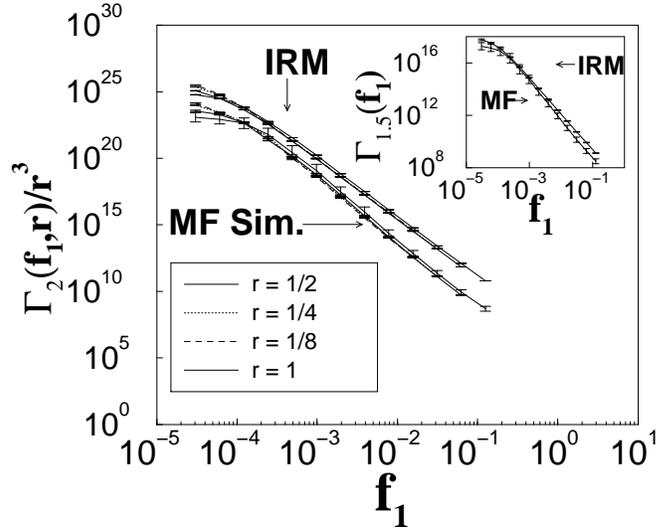}% Here is how to import EPS art
\caption{\label{haarsumplot} We present $\Gamma_2(f_1,r)$
($r=\frac{f_a}{f_b}=1$ corresponds to $\Gamma_2(f_1)$, and we set $f_1
= f_b$ for $r<1$) in the MF simulation and in the IRM.  $\Gamma_2(f_1,r)$
collapses for $r<1$, in excellent agreement with
Eq. (\ref{theta3SC}). For high frequency $\Gamma_2(f_1,r)\sim
f_1^{-Q_2}$, (see Table \ref{expo}). Inset: $\Gamma_{1.5}(f_1)$ for
the MF and the IRM. For high frequency $\Gamma_{1.5}(f_1)\sim
f_1^{-Q_{1.5}}$, see Table \ref{expo}.}
\end{figure}

The functions $\Gamma_{1.5}(f_1)$, $\Gamma_2(f_1)$, and
$\Gamma_2(f_b,r)$ are independent of $f_2$ since they resulted from
the $\theta=0$ evaluation of $\langle v(t,f_1)H(t+\theta,f_1)\rangle$,
$\langle H(t,f_1)H(t+\theta,f_1)\rangle$, and $\la
H(t_a+\theta,f_a)H(t_b,f_b)\ra$ (see Fig. \ref{haarsumplot}).  These
$f_2$ independent terms are referred to as Gaussian background terms
\cite{Seidler}. Now $\Gamma_{1.5}^{(N)}(f_1)$, $\Gamma_2^{(N)}(f_1)$,
and $\Gamma_2^{(N)}(f_b,r)$ are $\Gamma_{1.5}(f_1)$, $\Gamma_2(f_1)$,
and $\Gamma_2(f_b,r)$ normalized by $\langle H(t,f_1)\rangle_t$,
$\langle H(t,f_1)\rangle_t^2$, and $\la H(t_b,f_b)\ra_{t_b}\la
H(t_a,f_a) \ra_{t_a}$, respectively. The first terms of
Eq. (\ref{halfSC}) and Eq. (\ref{secondSC}) have no $f_1$ dependence,
since the $f_1$ dependence of these terms drops out after they are
normalized by $\langle H(t,f_1)\rangle_t$ and $\langle
H(t,f_1)\rangle_t^2$, respectively. Nevertheless, the lack of $f_1$
dependence in the first terms of second spectra and real $1.5$ spectra
is in excellent agreement with our simulation results, see
Fig. \ref{halfspecplot} and the inset of
Fig. \ref{secondspecplot}. Also, we defer the discussion of
$Im\{S_2(f_2,f_a,f_b)\}$ and $Im\{S_{1.5}(f_2,f_1)\}$, since they are
very sensitive to the non-stationary properties of the infinite
avalanche in the analytical model. Please refer to Appendix
\ref{calchigh} for details of how Eqs. (\ref{theta1}-\ref{crossSC})
are calculated.

We present our results in Eq. (\ref{theta1}-\ref{crossSC}) in terms of
general scaling forms since the mean field and IRM simulations, as
well as the experimental data, obey the same scaling form as mean
field theory, only with different exponents and non-universal
constants.

\section{SIMULATION}
\subsection{Mean Field Simulation}

        We perform $300$ runs of a simulation of the mean field RFIM.
        We collect data taken from $H \in [0,0.00125]$ at $R=0.79788$
        ($R_c=0.79788456$) in systems with $N = 15 \times 10^6$ spins,
        and $J=1$. From this data we determine our simulation results
        agree with the scaling forms given in
        Eq. (\ref{theta1}-\ref{crossSC}) with exponents given by Table
        \ref{expo}. See Fig. \ref{halfspecplot}-\ref{crosspecplot}.

\subsection{Infinite Range Model Simulation}

        We perform $60$ runs of a 3D simulation of the IRM.  The data
        was taken from $H \in [1.25,1.88]$ (from the slanted part of
        the hysteresis loop) at $R=2.2$ in system with $N = 400^3$
        spins, $J_{inf}=\frac{1}{N}$ and $J=1$.  Again results agree
        with Eq. (\ref{theta1}-\ref{crossSC}) with exponents given in
        Table \ref{expo}. Refer to
        Fig. \ref{halfspecplot}-\ref{crosspecplot}.

\subsection{Finite Size Effects} 

We study finite size effects in our simulation (mean field and IRM)
by examining higher order spectra for various system sizes.  We find
that for smaller system sizes the high frequency scaling (and
flattening due to the background term) is unchanged for second
spectra, real $1.5$ spectra, and real cross second spectra. However,
at low frequency the scaling regime of the second spectra, real $1.5$
spectra, and real cross second spectra rolls over (in the IRM and MF
simulations) at frequency $f_1 \simeq \frac{1}{T_{max}}$, where
$T_{max}$ is the maximum avalanche duration.  $T_{max}$ is system size
dependent where $T_{max}\sim L^z$, and $L = N^{1/d}$. In a $N=400^3$
system (IRM) we find that $T_{max}\simeq 4300$.

\begin{table}
\begin{tabular}{|c||c|c|c|c|c|}\hline
& $V_{1.5}$ & $V_2$ & $Q_{1.5}$ & $Q_2$ & $V_C$ \\ 
        \hline
        MFT & 2  & 2  & 3  &  5 & 2 \\
        \hline
        MF Sim. & $1.93\pm0.10$ & $1.92\pm0.12$ & $2.95\pm0.10$ & $4.93\pm0.12$
        & $1.92\pm0.15$\\
        \hline
        IRM ($d = 3$) & $1.80\pm0.07$ & $1.80\pm0.05$  & $2.73\pm0.06$ & $4.46\pm 0.07$ 
        & $1.82\pm 0.14$\\
        \hline
        Experiment & $0.93\pm0.20$ & $0.66\pm0.12$ & $2.81\pm0.08$(h.f.) & $4.39\pm0.15$(h.f.)
        & $0.63\pm0.06$\\
        & & & $1.66\pm0.12$(l.f) & $2.48\pm0.10$(l.f.) & \\
        \hline
\end{tabular}
%\vspace*{0.1in}
\caption{\label{expo} We present the values of the exponents:
$V_{1.5}$, $V_2$, $Q_{1.5}$, and $Q_2$ given in
Eqs. (\ref{theta1}-\ref{secondSC}) for mean field theory (MFT), mean
field simulation (MF Sim.), the infinite range model (IRM) for $d=3$,
and experiment. While the exponents values for MFT were determined
analytically, the exponents for the MF Sim., IRM, and experiment were
determined through a non-linear curve fitting of the data. In
particular, in our experimental plots
(Figs. \ref{haarsumexp}-\ref{crossexp}) we use the following window
sizes to fit the exponents: $f_1:[1kHz,200kHz]$ for $V_{1.5}$,
$f_1:[20Hz,2kHz]$ for $V_2$, $f_1:[14kHz,40kHz]$ for $Q_{1.5}$ (high
frequency exponent), $f_1:[.4kHz,18kHz]$ for $Q_2$ (high frequency
exponent), and $f_1:[20Hz,2kHz]$ for $V_C$. We find that that these
exponents do not change (outside of their error bars stated above)
when the window size for their measurement is changed within the
scaling regime of the data. Also, l.f. stands for low frequency and
h.f. stands for high frequency, since for the experiment there are
distinct l.f. and h.f. exponents, in some cases.}
\end{table}

\section{DISCUSSION}

The mean field theoretical calculation was for a single infinite
avalanche while the mean field simulation was obtained from a train of
avalanches.  Consequently, the train of avalanches introduces
intermittency that lowers the magnitude of the mean field exponents,
since the intermittency effectively adds white inter-avalanche noise
to the intra-avalanche noise seen in the MF calculation.  From Table
\ref{expo} we notice that the exponents for our mean field simulation
are systematically smaller (by an amount of $1\%$ to $4\%$) than the
exponents determined in mean field theory. Nevertheless, despite this
small deviate, our mean field simulation results agree very well with
mean field theory, corroborating our theoretical calculation.

The background components ($\theta=0$), given by $\Gamma_{1.5}(f_1)$ and
$\Gamma_2(f_b,r)$, are given in Fig.  \ref{haarsumplot} for mean field
theory and the mean field simulation.  For the corresponding exponents
$Q_{1.5}$ and $Q_2$ we find excellent agreement between mean field
theory and the mean field simulation results.  With the help of
\cite{Dahmen} we ascertain the following exponent relation for $Q_2$
and $Q_{1.5}$:

\begin{eqnarray} \label{scaling}
&& Q_2 = \frac{5-\tau}{\sigma\nu z} - 2\\
\label{scaling2}
&& Q_{1.5} = \frac{1}{\sigma\nu z} +1
\end{eqnarray}

Plugging $\tau$ and $1/\sigma\nu z$ (given in \cite{Mehta}) into Eqs.
(\ref{scaling}-\ref{scaling2}) we find exact agreement in mean field
theory: $Q_2 = 5$ and $Q_{1.5} = 3$. For the IRM in $d=3$ we find $Q_2
= 4.40\pm0.10$ and $Q_{1.5} = 2.72\pm0.03$, in close agreement with
the table above. Also, for experiment we find reasonable agreement
(within $10\%$) against the high frequency values for $Q_2$ and
$Q_{1.5}$; using Eqs. (\ref{scaling}-\ref{scaling2}) and \cite{Mehta}
we find $Q_2 = 2.70\pm0.05$ and $Q_{1.5} = 4.02\pm0.20$ for
experiment. Also, the measured experimental values for $Q_2$ and
$Q_{1.5}$ in the table above agree within error bars with the
exponents for the IRM.

\begin{figure}
\includegraphics[scale= .45]{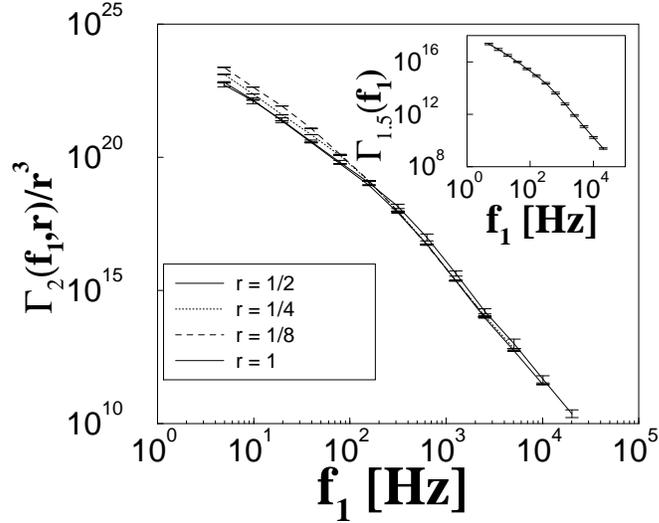}% Here is how to import EPS art
\caption{\label{haarsumexp} We present $\Gamma_2(f_1,r)$ (where
$r=\frac{f_a}{f_b}=1$ corresponds to $\Gamma_2(f_1)$, and we set
$f_1=f_b$ for $r<1$) in experiment. $\Gamma_2(f_1,r)$ collapses for
$r<1$, in excellent agreement with Eq. (\ref{theta3SC}).  For high
frequency $\Gamma_2(f_1,r)\sim f_1^{-Q_2}$, see Table
\ref{expo}. Inset: $\Gamma_{1.5}(f_1)$ for experiment. For high
frequency $\Gamma_{1.5}(f_1)\sim f_1^{-Q_{1.5}}$, see Table
\ref{expo}.}
\end{figure}

Interestingly, we notice that the background components for experiment (given
in Fig. \ref{haarsumexp}) have two scaling regimes: a flatter slope
at low frequency, a steeper slope at high frequency, and a transition
point at $f_{cross} \simeq 320$ Hz. This change in slope indicates that
intra-avalanche correlations are effecting the power at $f_1 <
f_{cross}$.

\begin{figure}
\includegraphics[scale= .45]{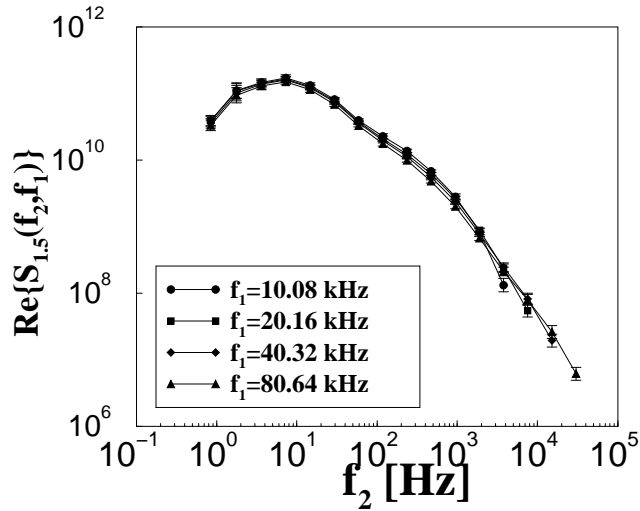}% Here is how to import EPS art
\caption{\label{halfspecexp} We present the real $1.5$ spectra
$Re\{S_{1.5}(f_2,f_1)\}$ in experiment.  At high frequency
$Re\{S_{1.5}(f_2,f_1)\}\sim f_2^{-V_{1.5}}$ (go to Table
\ref{expo}). The high frequency scaling regime is small for the
experimental real $1.5$ spectra since curve rolls over at $f_{cross} \simeq
320$ Hz.}
\end{figure}

\begin{figure}
\includegraphics[scale= .45]{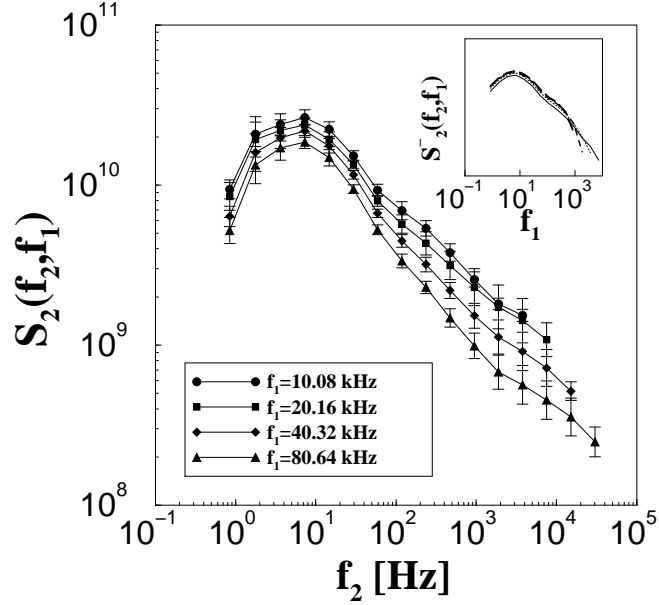}% Here is how to import EPS art
\caption{\label{secondexp}We present the second spectra,
$S_2(f_2,f_1)$, in experiment. At high frequency the scaling is given
$S_2(f_2,f_1)\sim f_2^{-V_2}$, where $V_2=0.66\pm0.12$, that is
significantly smaller than the mean field simulation
($V_2=1.92\pm0.12$) and the IRM ($V_2=1.80\pm0.05$). Further more,
there is no conspicuous flattening in these experimental second
spectra curves, indicating that the background term,
$\Gamma_2^{(N)}(f_1)$, is small for experiment versus the mean field
simulation and the IRM. Also, there is a noticeable separation between
curves at low frequency that is not present in the mean field
simulation or IRM second spectra curves. Inset: $S_2^-(f_2,f_1)$ is
the second spectra in experiment with the background term subtracted,
as a result the seperation between the curves vanishes, within error
bars. }
\end{figure}

The real $1.5$ spectra and the second spectra are given in Figs.
(\ref{halfspecplot}-\ref{secondspecplot},\ref{halfspecexp}-\ref{secondexp}).
Since the background term is small for the real $1.5$ spectra (for
IRM, MF simulation, and experiment) we find that the real $1.5$
spectra curves collapse upon themselves, in agreement with Eq.
(\ref{halfSC}).  The high frequency scaling exponent $V_{1.5}$ for the
real $1.5$ spectra shows excellent agreement between mean field theory
($V_{1.5} = 2$) and mean field simulation ($V_{1.5} = 1.93\pm0.10$).
The second spectra (Fig.  \ref{secondspecplot}) exhibits a conspicuous
flattening due to the background term, except in the experimental
second spectra (Fig.  \ref{secondexp}) where the flattening in
much less pronounced.  In order to find the high frequency scaling
exponent for the second spectra we do a non-linear curve fit using the
equation: $A0*x^{-A1}+$ (theoretical value, $\Gamma_2^{(N)}(f_1)$), where
$A0$, and $A1$, are free parameters.  We find very good agreement
between mean field theory ($V_2=2$) and our mean field
simulation ($V_2= 1.92\pm0.12$) for our second spectra exponent.

The very weak dependence of the real $1.5$ spectra on $f_1$ and the strong
fall off of the second spectra at high frequency suggests that the
high frequency power comes from the fine structure of large avalanches
($T>>1/f_1$, $T$ is the duration) and not small individual pulses
($T\simeq 1/f_1$)\cite{Petta}. Through simulation we verify this claim
in the mean field simulation and IRM. When we subtract all avalanches
smaller than $T=T_{max}/4$ from the Barkhausen train (where $T_{max}$
is the duration of the largest avalanche) and then determine the second
spectra, we find no change in $V_2$. However, when we subtract all
avalanches larger than $T=T_{max}/4$ from the Barkhausen train, we
find that the second spectrum flattens and that there is an evident
separation between the real $1.5$ spectra curves (i.e.  increased
$f_1$ dependence).  In experiment we also find a weak dependence of
the real $1.5$ spectra on $f_1$ and a strong fall off of the second
spectra, however, the second spectra for experiment falls off with an
exponent $V_2 = 0.66\pm0.12$ versus an exponent of $V_2 =
1.80\pm0.05$ for IRM and $V_2 = 2$ for mean field theory.  This
suggests that the high frequency power in experiment comes from the
fine structure of larger pulses to a {\em lesser degree} than in IRM
or mean field theory. This may be the result of dipole-dipole
interactions that are present in experiment. Since the dipole-dipole
interactions decay as a power law they may still be significant at
short length scales, thereby resulting in suppressed spin flips that
cause otherwise larger avalanches to be broken down in to smaller high
frequency pulses.  We are currently testing this hypothesis.

\begin{figure}
\includegraphics[scale= .45]{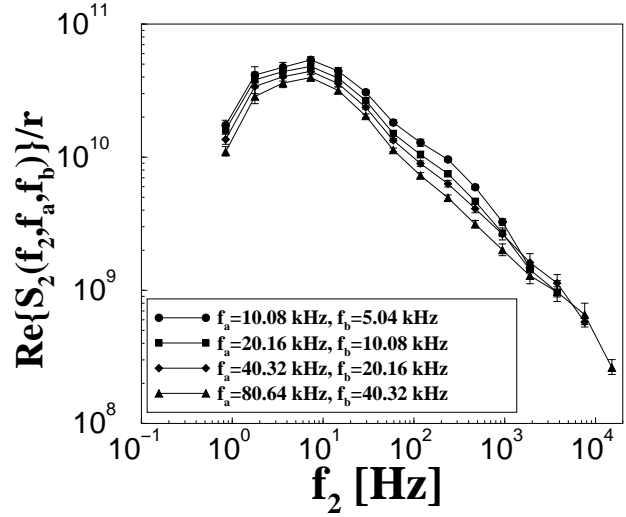}% Here is how to import EPS art
\caption{\label{crossexp}We present the real cross second spectra,
$Re\{S_2(f_2,f_a,f_b)\}$, for $r=f_a/f_b=\frac{1}{2}$, in
experiment. Again, as in the experimental second spectra there is no
conspicuous flattening due to the background term,
$\Gamma_2^{(N)}(f_b,r=\frac{1}{2})$. However, the seperation between
the curves vanishes, within error bars, when the background term is
subtrated off. }
\end{figure}

In Fig. \ref{crosspecplot} we give the real cross second spectra
($r=f_a/f_b=\frac{1}{2}$) for the mean field simulation and the IRM,
and in Fig. \ref{crossexp} we give the real cross second spectra for
experiment (also $r=\frac{1}{2}$). The cross second spectra plots for
$r<\frac{1}{2}$ are similar to $r=\frac{1}{2}$, so to avoid redundancy
$r<\frac{1}{2}$ plots have been left out of the paper. The real cross
second spectra not only strongly resemble the second spectra, but we
also notice in Table \ref{expo} that the exponents values for $V_2$
and $V_C$ (where $V_C$ was determined using the same non-linear curve
fit used to find $V_2$) in mean field theory, the mean field
simulation, the IRM, and in the experiment are identical or nearly
identical. For the class of models and systems we study in this paper
the real cross second spectra gives no new information, however, the
real cross second spectra is useful when studying systems that have
different dynamics on different length and time scales
\cite{Weissman}.

Remarkably, we have found that $Q_{1.5}$ and $Q_2$ agree (within error
bars) for the IRM and experiment. Since $Q_{1.5}$ and $Q_2$ are
directly related to known exponents, as we have found above, they may
be obtained from a standard analysis of power spectra ($P(w)\sim
w^{-\frac{1}{\sigma\nu z}}$) and the avalanche size distribution
($D(S)\sim S^{-\tau}$) \cite{Matt1}. However, when we compare
$V_{1.5}$, $V_2$, and $V_C$ between the IRM and experiment we find a
significant difference, whose origin we are currently
investigating. Thus by utilizing higher order spectra we present a
more rigorous test for avalanche models against experiment.

        Our study of higher order spectra is a powerful tool to
        further our understanding of noise in disordered systems. In
        addition, our mean field results are applicable to a large
        array of systems, in particular systems discussed in
        \cite{Fisher}.

\begin{acknowledgments}
  We would like to thank J. Sethna and T. Wotherspoon for helpful
  discussions, and we thank M. Kuntz and J. Carpenter for providing
  the front propagation model simulation code.  K.D. and
  A.P.M. acknowledge support from NSF via Grant Nos. DMR
  03-25939(ITR), the Materials Computation Center, through NSF Grant
  No. 03-14279, and IBM which provided the computers that made the
  simulation work possible. M.B.W. and A.C.M.  acknowledge support
  from NSF via Grant No. DMR 02-40644.  A.P.M.  would also like to
  acknowledge the support provided by UIUC through a University
  Fellowship, and K.D. gratefully acknowledges support through an
  A.P. Sloan fellowship.
\end{acknowledgments}

\begin{appendix}

\section{ Calculation of Correlation functions }
\label{corcalc}
        From Eq. (\ref{prodist}) one can verify the following recursion relations:

\begin{eqnarray}
<n_j>&=&<n_{j-1}>\\
<n_j^2>&=&<n_{j-1}^2>+<n_{j-1}>\\
<n_j^3>&=&<n_{j-1}^3>+3<n_{j-1}^2>+<n_{j-1}>\\
<n_j^4>&=&<n_{j-1}^4>+6<n_{j-1}^3>+7<n_{j-1}^2>+<n_{j-1}>
\end{eqnarray}

        Using the fact that $n_o \equiv 1$ we determine the following
        initial conditions:

\begin{eqnarray}
<n_1>&=&1\\
<n_1^2>&=&2\\
<n_1^3>&=&5\\
<n_1^4>&=&15
\end{eqnarray}

        Using the recursion relations and initial conditions we can find 
        the explicit functionality of $<n_j^2>$, $<n_j^3>$, and $<n_j^4>$:

\begin{eqnarray*}
<n_j^2>&=&<n_{j-1}^2>+<n_{j-1}>\\
&=&<n_{j-2}^2>+<n_{j-2}>+<n_{j-1}>\\
\ldots &=&<n_1^2>+\sum_{m=1}^{j-1}<n_{j-m}>\\
&=&2+\sum_{m=1}^{j-1}1\\
&=&j+1
\end{eqnarray*}

        Thus we verify Eq. (\ref{avg3}):

\begin{equation}
<n_jn_{j+k}> = <n_j^2> = j+1
\end{equation}

        Now let us consider:

\begin{eqnarray*}
<n_j^3>&=&<n_{j-1}^3>+3<n_{j-1}^2>+<n_{j-1}>\\
&=&<n_{j-2}^3>+3(<n_{j-2}^2>+<n_{j-1}^2>)+<n_{j-1}>+<n_{j-2}>\\
\ldots &=&<n_1^3>+3\summ<n_{j-m}^2>+\summ<n_{j-m}>\\
&=&5+3\summ (j-m+1)+\summ 1\\
&=& \frac{1}{2}(3j^2+5j+2)
\end{eqnarray*}

        Thus we find:

\be
<n_j^3>=\frac{1}{2}(3j^2+5j+2)
\ee

        Now let us find $<n_j^4>$:

\bea
<n_j^4>&=&<n_{j-1}^4>+6<n_{j-1}^3>+7<n_{j-1}^2>+<n_{j-1}>\\
&=& 15 +\summ\lbrack6<n_{j-m}^3>+7<n_{j-m}^2>+<n_{j-m}>\rbrack\\
&=& \frac{1}{2}(6j^3+13j^2+9j+2)
\eea

        So we have:

\be
<n_j^4> = \frac{1}{2}(6j^3+13j^2+9j+2)
\ee

        Now let us look at the case 
        where $l\ge k\ge 0$:

\bea
<n_jn_{j+k}n_{j+l}> &=& <n_in_{j+k}^2> \\
&=& <n_j(n_{j+k-1}+n_{j+k-1}^2)>\\
&=& <n_j^2>+<n_jn_{j+k-1}^2>\\
&=& 2<n_j^2>+<n_jn_{j+k-2}^2>\\
&& \ldots \\
&=& k<n_j^2>+<n_j^3>
\eea

        Thus:

\be
<n_jn_{j+k}n_{j+l}> = <n_j^3>+k<n_j^2>
\ee

        Now let's look at the most general situation where
        $m\ge l\ge k\ge 0$:

\bea
<n_jn_{j+k}n_{j+l}n_{j+m}> &=& <n_jn_{j+k}n_{j+l}^2>\\
&=& <n_jn_{j+k}(n_{j+l-1}^2+n_{j+l-1})>\\
&=& <n_jn_{j+k}^2>+<n_jn_{j+k}n_{j+l-1}^2>\\
&=& <n_jn_{j+k}^2>+<n_jn_{j+k}(n_{j+l-2}^2+n_{j-l-2})>\\
&=& 2<n_jn_{j+k}^2>+<n_jn_{j+k}n_{j+l-2}^2>\\
&& \ldots\\
&=& (l-k)<n_jn_{j+k}^2> + <n_jn_{j+k}^3>\\
&=& (l-k)\lbrack<n_j^3>+k<n_j^2>\rbrack + <n_jn_{j+k}^3>
\eea

        Now let's determine $<n_jn_{j+k}^3>$:

\bea
<n_jn_{j+k}^3>&=&<n_j(n_{j+k-1}^3+3n_{j+k-1}^2+n_{j+k-1})>\\
&=&<n_jn_{j+k-1}^3>+3<n_jn_{j+k-1}^2>+<n_jn_{j+k-1}>\\
&& \ldots\\
&=&<n_j^4>+3\sum_{m=1}^{k-1}<n_jn_{j+m}^2>+k<n_j^2>\\
&=&<n_j^4>+3\sum_{m=1}^{k-1}\lbrack<n_j^3>+m<n_j^2>\rbrack+k<n_j^2>\\
&=&<n_j^4>+3k<n_j^3> + \frac{k}{2}(3k-1)<n_j^2>
\eea

        Thus combining the above results we verify Eq. (\ref{avg6}):
\be
<n_jn_{j+k}n_{j+l}n_{j+m}> = <n_j^4>+(2k+l)<n_j^3>+\frac{k}{2}(k+2l-1)<n_j^2> 
\ee

\section{ Time Ordered products }
\label{timeorder}

        In order to evaluate Eq. (\ref{haar1}) we need to consider
        the time ordering of the indices of $n$, consider the following
        ($i<N$):

\begin{eqnarray}\label{example}
        && \tilde{y} = n_1 + n_2 +\ldots+n_{N-1}+n_N\\
        && x_i = n_i+n_{i+1}+\ldots+n_{N-1}+n_N
\end{eqnarray}

        Now we want to evaluate $\tilde{y}^2$, $\tilde{y}^3$, $\tilde{y}^4$
        we must write the expansion as a sum of time-ordered products,
        here's how we do it:

\begin{eqnarray}\label{induct}
        && \tilde{y}^2 = (n_1+x_2)^2 = n_1^2 + 2n_1x_2 + x_2^2 \\
        && x_2^2 = (n_2+x_3)^2 = n_2^2 + 2n_2x_3 + x_3^2\\
        && x_3^2 = (n_3+x_4)^2 = n_3^2 + 2n_3x_4 + x_4^2\\
        && \ldots
\end{eqnarray}
        
        Thus by we can write Eq. (\ref{induct}) as:

\begin{eqnarray}
\tilde{y}^2 &=&\sum_{i=1}^N \lbrack n_i^2 + 2n_ix_{i+1}\rbrack\\
\label{order}
&=&\sum_{i=1}^N \lbrack n_i^2 + 2n_i\sum_{j=i+1}^Nn_j\rbrack\\
\end{eqnarray}

        So we see that we have successfully written $\tilde{y}^2$ as a
        sum of time-ordered products.  Now we can easily evaluate
        $<\tilde{y}^2>$ using Eqs. (\ref{avg2})-(\ref{avg3}).  This
        remarkable method can be used for higher powers to
        evaluate $\tilde{y}^3$, and $\tilde{y}^4$.

\section{Calculation of Higher Order Spectra}
\label{calchigh}
\subsection{ Second Spectra }

We first rewrite the second spectra as the follows:

\bea
&& S_2(\omega,N) = \frac{\langle
F_i\{H(i,N)\}F_i\{H(i,N)\}\rangle}{\langle H(i,N)\rangle_i^2} \\
&&=
\frac{\sum\limits_{\theta=0}^{T/N-1} e^{2\pi i\theta \omega N/T}
\sum\limits_{i=0}^{T/N-1-\theta} \langle
H(i,N)H(i+\theta,N)\rangle}{\langle H(i,N)\rangle_i^2} \\
\eea

Where $i = t/N$, $\langle...\rangle_i$ denotes sum over $i$
($\sum_{i=0}^{T/N-1}$), and $F_i\{..\}$ represents the discrete
Fourier transform over $i$. The variable $\omega$ is conjugate
variable to $i$ in the Fourier transform; $f_2 = \omega/N$ is
conjugate to $t$ (the original time) since $t = Ni$. We will use
$\omega$ in subsequent calculations for consistency and do a change of
variable to $f_2$ in our final expression. Also, we have set $\Delta t
= 1$.

        Let's first consider the second spectra, when $\theta>0$ we
        have:

\begin{equation} \label{secondspec3}
\langle H(i+\theta,N)H(i,N) \rangle 
= \langle (\tilde{x}_{i+\theta}^+(N)-\tilde{x}_{i+\theta}^-(N))^2 (\tilde{x}_i^+(N)-\tilde{x}_i^-(N))^2 \rangle
\end{equation}

        This product is already partially time ordered since we know
        $\theta>0$, however when $\theta=0$:

\begin{equation} \label{secondspec4}
\langle H(i,N)^2 \rangle 
= \langle (\tilde{x}_i^+(N)-\tilde{x}_i^-(N))^4 \rangle
\end{equation}

        In this case since $\theta=0$ no simplification can be made,
        there is no partial time ordering like in the case of
        Eq. (\ref{secondspec3}).
        
        We begin by evaluating Eq. (\ref{secondspec4}), which can be
        written as a sum of 4-pt functions (Eq.(\ref{avg6})).  To write
        Eq. (\ref{secondspec4}) as a sum of 4-pt functions we must use
        the method discussed in Appendix \ref{timeorder}, which allows
        us to write any power of $\tilde{x}_i^{\pm}(N)$ as a sum of
        time ordered products in $n$:

\begin{eqnarray*}
&&\langle H(i,N)^2 \rangle \\
&&= \langle (\tilde{x}_i^+-\tilde{x}_i^-)^4 \rangle \\ 
&&=\langle (\tilde{x}_i^+)^4-4(\tilde{x}_i^+)^3\tilde{x}_i^- 
+ 6(\tilde{x}_i^+)^2(\tilde{x}_i^-)^2-4\tilde{x}_i^+(\tilde{x}_i^-)^3+(\tilde{x}_i^-)^4\rangle \\
&&=\sum_{j=0}^{N/2-1} \Bigg[\Big[4(N/2-1-j)+1\Big]\langle n_{iN+j}^4\rangle 
+ 6\sum_{k=j+1}^{N/2-1}\Big[2(N/2-1-k)+1\Big]\langle n_{iN+j}^2 n_{iN+k}^2\rangle  \\
&&+ 4\sum_{k=j+1}^{N/2-1}\Big[(3(N/2-1-k)+1)\langle n_{iN+j}n_{iN+k}^3\rangle
+ 3\sum_{l=k+1}^{N/2-1}(2(N/2-1-l)+1)\langle n_{iN+j}n_{iN+k}n_{iN+l}^2\rangle\Big]\Bigg] \\
&& - 2N\sum_{j=0}^{N-1}\Bigg[\langle n_{iN+j}^4\rangle 
+ \sum_{l=j+1}^{N/2-1}\Big[3\langle n_{iN+j}^2 n_{iN+l}^2\rangle 
+ 3\langle n_{iN+j}n_{iN+l}^3\rangle 
+ 6\sum_{m=l+1}^{N/2-1}\langle n_{iN+j}n_{iN+l}n_{iN+m}^2\rangle\Big]\Bigg] \\ 
&&+ 6\sum_{j=0}^{N/2-1}\sum_{k=N/2}^{N-1}\Big[2(N-1-k)+1\Big]\Big[\langle n_{iN+j}^2 n_{iN+k}^2\rangle 
+ 2\sum_{l=j+1}^{N/2-1}\langle n_{iN+j}n_{iN+l}n_{iN+k}^2\rangle\Big] \\ 
&&- 4\sum_{j=0}^{N/2-1}\sum_{k=N/2}^{N-1}\Bigg[(3(N-1-k)+1)\langle n_{iN+j}n_{iN+k}^3\rangle 
+ 3\sum_{l=k+1}^{N-1}(2(N-1-l)+1)\langle n_{iN+j}n_{iN+k}n_{iN+l}^2\rangle\Bigg] \\
&&+ \sum_{j=N/2}^{N-1}\Bigg[\Big[4(N/2-1-j)+1\Big]\langle n_{iN+j}^4\rangle 
+ 6\sum_{k=j+1}^{N-1}\Big[2(N/2-1-k)+1\Big]\langle n_{iN+j}^2 n_{iN+k}^2\rangle  \\
&&+ 4\sum_{k=j+1}^{N-1}\Big[(3(N-1-k)+1)\langle n_{iN+j}n_{iN+k}^3\rangle 
+ 3\sum_{l=k+1}^{N-1}(2(N-1-l)+1)\langle n_{iN+j}n_{iN+k}n_{iN+l}^2\rangle\Big]\Bigg] 
\end{eqnarray*}

        Performing the above sum we find:

\begin{eqnarray}
&&\langle H(i,N)^2 \rangle \\
&&=\frac{N\left[64+56(21-20i)N^2-420N^3 + 14(58+80i)N^4-105N^5+4(61+70i)N^6\right]}{13440}
\end{eqnarray}

        Finally summing over $i$ we obtain:

\begin{eqnarray}
&&\Gamma_2(N) = \sum_{i=0}^{T/N-1}\langle H(i,N)^2 \rangle \\
&&= \frac{1}{96}N(N^2+2)^2 T^2 + O(T) \\
&&\simeq \frac{1}{96}N^5T^2
\end{eqnarray}

        Now let's consider the case where $\theta > 0$:

\begin{eqnarray*}
&& \langle H(i+\theta,N)H(i,N) \rangle \\ &&= \langle
(\tilde{x}_i^+(N)-\tilde{x}_i^-(N))^2(\tilde{x}_{i+\theta}^+(N)
-\tilde{x}_{i+\theta}^-(N))^2 \rangle \\ &&=
\Bigg[\sum_{l=0}^{N/2-1}\Big[2(N/2-1-l)+1-N\Big] +
\sum_{l=N/2}^{N-1}\Big[2(N-1-l)+1\Big]\Bigg]\times \\
&&\Bigg[\sum_{j=0}^{N/2-1}\Big[2\sum_{k=j+1}^{N/2-1} \la
n_{iN+j}n_{iN+k}n_{(i+\theta)N+l}^2\ra + \la
n_{iN+j}^2n_{(i+\theta)N+l}^2\ra\Big]
-2\sum_{j=0}^{N/2-1}\sum_{k=N/2}^{N-1} \la
n_{iN+j}n_{iN+k}n_{(i+\theta)N+l}^2\ra \\ && +
\sum_{j=N/2}^{N-1}\Big[2\sum_{k=j+1}^{N-1} \la
n_{iN+j}n_{iN+k}n_{(i+\theta)N+l}^2\ra +\la
n_{iN+j}^2n_{(i+\theta)N+l}^2\ra\Big]\Bigg]\\ &&=
\frac{1}{576}N^3(N^2+2)(5N^2+4+4i(N^2+2))
\end{eqnarray*}

        After summing over $i$ and performing Fourier transform we find:

\begin{eqnarray*}
&&\sum_{\theta=1}^{T/N-1} e^{2\pi i\theta \omega/T}
\sum_{t=0}^{T/N-1-\theta} \langle H(t+\theta,N)H(t,N)\rangle \\
&& = \frac{N^2(N^2+2)^2T}{144\omega^2}
\end{eqnarray*}

        From Eq.(\ref{haar2}) we know $\langle H(i,N)\rangle_t^2 = \frac{T^2}{144}(N^2+2)^2$,
        also $f_2=\omega/N$, and $f_1 = 1/N$, so finally we obtain: 

\begin{equation}
S_2(f_2,f_1) \simeq  \frac{1}{Tf_2^2}+ \frac{2}{3f_1}
\end{equation}

\subsection{ $1.5$ Spectra }

We first rewrite the $1.5$ spectra equation in a form we may
analytically evaluate:

\bea
&& S_{1.5}(\omega,N) = \frac{\langle
F_i\{v(i,N)\}F_i\{H(i,N)\}\rangle}{\langle H(i,N)\rangle_i} \\
&&= \frac
{ \sum\limits_{\theta=0}^{T/N-1} e^{2\pi i\theta \omega N/T}
\sum\limits_{i=0}^{T/N-1-\theta} \langle
v(i,N)H(i+\theta,N)\rangle}{\langle H(i,N)\rangle_i}\\
\eea

        Where $v(i,N) \equiv \tilde{x}_i^+(N)+\tilde{x}_i^-(N)$.
        Now we consider the $1.5$ spectra, when $\theta=0$ we
        have:

\begin{eqnarray*} \label{halfspec2}
&& \Gamma_{1.5}(N) = \sum_{i=0}^{T/N-1}\langle v(t,N)H(i,N) \rangle 
= \Big[\sum_{l=0}^{N/2-1}n_{iN+l}+\sum_{l=N/2}^{N-1}n_{iN+l}\Big](\tilde{x}_i^+(N)-\tilde{x}_i^-(N))^2 \\ 
&& = \sum_{i=0}^{T/N-1}\Bigg[\sum_{j=0}^{N/2-1}\Big[(3(N/2-1-j)+1)\langle n_{iN+j}^3\rangle
 + \sum_{k=j+1}^{N/2-1}3(2(N/2-1-k)+1)\langle n_{iN+j}^2n_{iN+k}\rangle\Big] \\
&& - N/2\sum_{j=0}^{N/2-1}\Big[\langle n_{iN+j}^3\rangle + 
2\sum_{k=j+1}^{N/2-1}\langle n_{iN+j}n_{iN+k}^2\rangle\Big]
 -\sum_{j=0}^{N/2-1}\sum_{k=N/2}^{N-1}\Big[2(N-1-k)\langle n_{iN+j}n_{iN+k}^2\rangle\Big]\Bigg]\\
&& = \frac{1}{24}N(N^2+2)T^2+O(T)\\
&& \simeq \frac{1}{24}N^3T^2
\end{eqnarray*}

        For the case $\theta>0$ we find:

\begin{eqnarray*} \label{halfspec3}
&& \langle v(i,N)H(i+\theta,N) \rangle 
= \sum_{l=0}^{N-1}n_{iN+l}(\tilde{x}_{i+\theta}^+(N)-\tilde{x}_{i+\theta}^-(N))^2 \\ 
&& = \sum_{l=0}^{N-1}\Bigg[\sum_{j=0}^{N/2-1}
\Big[(2(N/2-1-j)+1-N\Big]\langle n_{iN+l}n_{(i+\theta)N+j}^2\rangle\\ 
&& +\sum_{j=N/2}^{N-1}\Big[2(N-1-j)+1\Big]\langle n_{iN+l}n_{(i+\theta)N+j}^2\rangle\Bigg]\\
&& = \frac{1}{24}N(N^2+2)(2Ni+N+1)
\end{eqnarray*}

        After summing over $i$, performing Fourier transform, and
        taking the real part we find:

\begin{eqnarray*}
&&Re\{\sum_{\theta=1}^{T/N-1} e^{2\pi i\theta \omega/T}
\sum_{i=0}^{T/N-1-\theta} \langle v(i,N)H(i+\theta,N)\rangle\} \\
&& = \frac{N^2(N^2+2)T}{12\omega^2}
\end{eqnarray*}

        Now we normalize the result with $\langle H(i,N)\rangle_t =
        \frac{T}{12}(N^2+2)$, and substitute $f_2=\omega/N$, and $f_1
        = 1/N$.

\begin{equation}
Re\{S_{1.5}(f_2,f_1)\} = \frac{1}{f_2^2} + \frac{f_1}{2}T
\end{equation}

\subsection{ Cross Second Spectra }

We rewrite the cross second spectra as follows:

\bea
&& S_2(\omega,M,N) = \frac{\langle
F_j\{(j,M)\}F_i\{H(i,N)\}\rangle}{\langle H(i,N)\rangle_i\langle
H(j,M)\rangle_j}\\
&& = \frac{\sum\limits_{\theta=0}^{T/N-1} e^{2\pi
i\theta \omega N/T} \sum\limits_{i=0}^{T/N-1-\theta}\frac{M}{N}
\sum\limits_{j=N(i-1)/M+1}^{Ni/M} \langle
H(j,M)H(i+\theta,N)\rangle}{\langle H(i,N)\rangle_i\langle
H(j,M)\rangle_j}
\eea

Where $M=\frac{1}{f_b}$, $N=\frac{1}{f_a}$, and $\langle...\rangle_j$
denotes sum over $j$ ($\sum_{j=0}^{T/M-1}$), Before we calculation of
the cross second spectra we find it useful to define the following
notation:

\begin{eqnarray*}
&& \ym_j(M) = \sum_{l=M/2}^{M-1} n_{Mj+l} \\
&& \yp_j(M) = \sum_{l=0}^{M/2-1} n_{Mj+l} \\
&& a = Ni/M \\
&& b = N(2i+1)/(2M)-1 \\
&& c = N(i+1)/M - 1 \\
&& \xm_i(N) = \sum_{k=b+1}^c (\yp_k(M)+\ym_k(M))=\sum_{j=N/2}^{N-1}n_{Ni+j}\\
&& \xp_i(N) = \sum_{k=a}^b (\yp_k(M)+\ym_k(M))=\sum_{j=0}^{N/2-1}n_{Ni+j}\\
&& Y_k = \yp_k(M) + \ym_k(M) \\
&& X_j = \yp_j(M) - \ym_j(M) 
\end{eqnarray*}

        Again first we consider case where $\theta=0$, using the
        notation above we have:

\begin{eqnarray}
&&\Gamma_2(N,r) = \sum\limits_{i=0}^{T/N-1}\frac{M}{N}
\sum\limits_{j=a}^c \langle H(j,M)H(i,N)\rangle\\ 
&& =\sum\limits_{i=0}^{T/N-1}\frac{M}{N} \sum\limits_{j=a}^c 
\langle (\tilde{y}_j^+(M)-\tilde{y}_j^-(M))^2 
(\tilde{x}_i^+(N)-\tilde{x}_i^-(N))^2 \rangle \\
&& = \frac{M}{N}\sum\limits_{i=0}^{T/N-1}\sum\limits_{j=a}^c
X_j^2\Big[\sum\limits_{j=a}^b Y_k-\sum\limits_{j=b+1}^c Y_k\Big]^2\\
\label{gamma_r}
&& = \frac{M}{N}\sum\limits_{i=0}^{T/N-1}\Big[\sum\limits_{j=a}^b
X_j^2 + \sum\limits_{j=b+1}^c X_j^2\Big]\Big[\sum\limits_{j=a}^b 
Y_k-\sum\limits_{j=b+1}^c Y_k\Big]^2
\end{eqnarray}
        
        Now we make the follow approximation for Eq. (\ref{gamma_r}),
        we simplify the limits by setting $N/M = 2$ in the limits.  By
        deduction this approximation will preserve the scaling for any
        $N/M$ (multiples of 2) in our final answer.  Using $N/M=2$ to
        simplify the limits we find:

\begin{eqnarray*}
&& a = 2i \\
&& b = 2i \\
&& c = b+1 = 2i+1 \\
\end{eqnarray*}
        
        We now expand Eq. (\ref{gamma_r}):

\begin{eqnarray*}
&& \Gamma_2(N) = \frac{M}{N}\sum\limits_{i=0}^{T/N-1}
\Big(\xt^2+\xtp^2\Big)\Big(\yt-\ytp\Big)^2 \\ 
&& = \frac{M}{N}\sum\limits_{i=0}^{T/N-1}
\Big(\xt^2\yt^2 + \xtp^2\ytp^2 + \xtp^2\yt^2 + \xt^2\ytp^2
-2\xt^2\yt\ytp-2\xtp^2\ytp\yt\Big)
\end{eqnarray*}

        We evaluate the above terms below as a function of $k$,
        where $k$ represents either $2i$ or $2i+1$:

\begin{eqnarray*}
&& X_{k+1}^2Y_{k+1}Y_k \\
&& = \sum_{m=0}^{M-1}\qful\Bigg[\Big[3(M-1-q)+1\Big]\langle n_{kM+m}n_{(k+1)M+q}^3\rangle\\
&&+ 3 \sum_{l=q+1}^{M-1}\Big[2(M-1-l)+1\Big]\langle n_{kM+m}n_{(k+1)M+q}n_{(k+1)M+l}^2 \rangle\Bigg]\\
&&-\sum_{m=0}^{M-1}\lful\qhaf\Bigg[\Big[2(M-1-l)+1\Big]
\langle n_{kM+m}n_{(k+1)M+q}n_{(k+1)M+l}^2\rangle\Bigg]\\
&&-M\sum_{m=0}^{M-1}\lhaf\Big[\langle n_{kM+m}n_{(k+1)M+l}^3\rangle 
+ 2\sum_{q=l+1}^{M/2-1}\langle n_{kM+m}n_{(k+1)M+l}n_{(k+1)M+q}^2\rangle\Big]\\
&& + \sum_{m=0}^{M-1}\qhaf\Bigg[\Big[3(M/2-1-q)+1\Big]\langle n_{kM+m}n_{(k+1)M+q}^3\rangle\\
&&+ 3 \sum_{l=q+1}^{M/2-1}\Big[2(M/2-1-l)+1\Big]\langle n_{kM+m}n_{(k+1)M+q}n_{(k+1)M+l}^2 \rangle\Bigg]\\
\end{eqnarray*}

\begin{eqnarray*}
&& X_{k+1}^2Y_k^2\\
&& = \Bigg[\mhaf\Big[2(M/2-1-m)+1-M\Big]+\mful\Big[2(M-1-m)+1\Big]\Bigg]\\
&&\times\Big[\sum_{n=0}^{M-1}\Big(\langle n_{kM+n}^2n_{(k+1)M+m}^2\rangle
+2\sum_{l=n+1}^{M-1}\langle n_{kM+n}n_{kM+l}n_{(k+1)M+n}^2\rangle\Big)\Big]
\end{eqnarray*}

\begin{eqnarray*}
&& X_k^2Y_k^2 \\
&& = \jful\Bigg[\Big[4(M-1-j)+1\Big]\langle n_{Mk+j}^4\rangle\\
&&+6\sum_{m=j+1}^{M-1}\Big[2(M-1-m)+1\Big]\langle n_{Mk+j}^2n_{Mk+m}^2 \rangle\\
&&+\sum_{m=j+1}^{M-1}4\Big[\Big(3(M-1-m)+1\Big)\langle n_{Mk+j}n_{Mk+m}^3\rangle\\
&&+\sum_{l=m+1}^{M-1}3\Big(2(M-1-l)+1\Big)\langle n_{Mk+j}n_{Mk+m}n_{Mk+l}^2\rangle\Big]\Bigg]\\
&& -2\jhaf\mful\Big[2(M-1-m)+1\Big]\times\Big[\langle n_{Mk+j}^2n_{Mk+m}^2\rangle + 
\sum_{l=j+1}^{M/2-1}2\langle n_{Mk+j}n_{Mk+l}n_{Mk+m}^2\rangle\Big]+\\
&& \jhaf\Bigg[\Big[4(M/2-1-j)+1\Big]\langle n_{Mk+j}^4\rangle\\
&&+6\sum_{m=j+1}^{M/2-1}\Big[2(M/2-1-m)+1\Big]\langle n_{Mk+j}^2n_{Mk+m}^2 \rangle\\
&&+\sum_{m=j+1}^{M/2-1}4\Big[\Big(3(M/2-1-m)+1\Big)\langle n_{Mk+j}n_{Mk+m}^3\rangle\\
&&+\sum_{l=m+1}^{M/2-1}3\Big(2(M/2-1-l)+1\Big)\langle n_{Mk+j}n_{Mk+m}n_{Mk+l}^2\rangle\Big]\Bigg]\\
\end{eqnarray*}

\begin{eqnarray*}
&& X_k^2Y_kY_{k+1}\\
&& = M\mful\Bigg[\la n_{kM+m}^4 \ra + 3\sum_{l=m+1}^{M-1}\la n_{kM+m}^2n_{kM+l}^2\ra\\
&& + 3\sum_{l=m+1}^{M-1}\Big[\la n_{kM+m}n_{kM+l}^3\ra 
+ 2\sum_{q=l+1}^{M-1}\la n_{kM+m}n_{kM+l}n_{kM+q}^2\ra\Big]\Bigg]\\
&&-M\mhaf\lful\Big[\la n_{kM+m}n_{kM+l}^3\ra+2\sum_{q=l+1}^{M-1}\la n_{kM+m}n_{kM+l}n_{kM+q}^2\ra\Big]\\
&&-M\mhaf\lful\Big[\la n_{kM+m}^2n_{kM+l}^2\ra+2\sum_{q=m+1}^{M-1}\la n_{kM+m}n_{kM+q}n_{kM+l}^2\ra\Big]\\
&& + M\mhaf\Bigg[\la n_{kM+m}^4 \ra + 3\sum_{l=m+1}^{M/2-1}\la n_{kM+m}^2n_{kM+l}^2\ra\\
&& + 3\sum_{l=m+1}^{M/2-1}\Big[\la n_{kM+m}n_{kM+l}^3\ra 
+ 2\sum_{q=l+1}^{M/2-1}\la n_{kM+m}n_{kM+l}n_{kM+q}^2\ra\Big]\Bigg]\\
\end{eqnarray*}

\begin{eqnarray*}
&& X_k^2Y_{k+1}^2 = \sum_{n=0}^{M-1}\Big[2(M-1-n)+1\Big]\\
&&\times\Bigg[\sum_{n=0}^{M-1}\mful\Big(\la n_{kM+m}^2n_{(k+1)M+n}^2\ra+2\sum_{l=m+1}^{M-1}
\la n_{kM+m}n_{kM+l}n_{(k+1)M+n}^2 \ra \Big)\\
&&-2\mhaf\lful\la n_{kM+m}n_{kM+l}n_{(k+1)M+n}^2\ra+\\
&& \sum_{m=0}^{M/2-1}\Big(\la n_{kM+m}^2n_{(k+1)M+n}^2\ra
+ 2\sum_{l=m+1}^{M-1}\la n_{kM+m}n_{kM+l}n_{(k+1)M+n}^2\ra\Big)\Bigg]
\end{eqnarray*}

        Now we can evaluate Eq. (\ref{gamma_r}) (where $r=M/N$):

\be
\Gamma_2(N,r) = \frac{M^5r^3}{144}\Big(5+\frac{4}{M^3}\Big)T^2 \simeq \frac{25M^5r^3}{144}T^2
\ee

        The $\theta > 0$ case is readily evaluated as followed:

\begin{eqnarray*}
&& \la H(j,M)H(i+\theta,N)\ra = \la (\yp_j-\ym_j)^2(\xp_{i+\theta}-\xm_{i+\theta})^2\ra\\
&& = \Bigg[\lhaf\Big[2(N/2-1-l)+1-N\Big]+\lful\Big[2(N-1-l)+1\Big]\Bigg]\\
&& \times\Bigg[\mhaf\Big[2\sum_{k=m+1}^{M/2-1}\la n_{jM+m}n_{jM+k}n_{(i+\theta)N+l}^2\ra 
+ \la n_{jM+m}^2n_{(i+\theta)N+l}^2\ra\Big] \\
&& -2\mhaf\kful\la n_{jM+n}n_{jM+k}n_{(i+\theta)N+l}^2\ra\\
&& +\mful\Big[2\sum_{k=m+1}^{M-1}\la n_{jM+m}n_{jM+k}n_{(i+\theta)N+l}^2\ra
+ \la n_{jM+m}^2n_{(i+\theta)N+l}^2\ra\Big]\Bigg]
\end{eqnarray*}

        After summing over $i$ and performing Fourier transform and
        taking the real part we find:

\begin{eqnarray*}
&&Re\{\sum_{\theta=1}^{T/N-1} e^{2\pi i\theta \omega/T}
\sum_{i=0}^{T/N-1-\theta} \langle H(j,M)H(i+\theta,N)\rangle\} \\
&& = \frac{MN(M^2+2)(N^2+2)T}{144\omega^2}
\end{eqnarray*}

        From Eq.(\ref{haar2}) we know $\la H(i,N)\ra_t \la H(j,M)\ra_t
        = \frac{T^2}{144}(N^2+2)(M^2+2)$, $f_2 = \omega/N$, $f_a =
        1/N$, and $f_b = 1/M$ finally we obtain:

\be
Re\{S_2(f_2,f_a,f_b)\}\simeq\frac{r}{Tf_2^2}+ \frac{25 r^5}{f_a}
\ee

\end{appendix}

\break

\end{document}